\documentclass[preprints,article,accept,moreauthors,pdftex]{Definitions/mdpi}
\firstpage{1} 
\pubvolume{1}
\issuenum{1}
\articlenumber{0}
\pubyear{2022}
\copyrightyear{2022}
\datereceived{} 
\dateaccepted{} 
\datepublished{} 
\hreflink{https://doi.org/} 

\usepackage{subcaption}

\Title{Electron-nucleus scattering in the NEUT event generator}

\TitleCitation{Electron-nucleus scattering in the NEUT event generator}

\Author{Stephen Dolan$^{1}$, Jordan McElwee$^{2}$, Sara Bolognesi$^{3}$, Yoshinari Hayato$^{4}$, Kevin McFarland$^{5}$, Guillermo Megias$^{6}$, Kajetan Niewczas$^{7,8}$, Luke Pickering$^{9}$, Jan Sobczyk$^{7}$, Lee Thompson$^{2}$, and Clarence Wret$^{5}$}

\AuthorNames{Stephen Dolan, Jordan McElwee, Sara Bolognesi, Yoshinari Hayato, Kevin McFarland, Guillermo Megias, Kajetan Niewczas, Luke Pickering, Jan Sobczyk, Lee Thompson, and Clarence Wret}

\AuthorCitation{Dolan, S.; McElwee, J.; Bolognesi, S.; Hayato, Y.; McFarland, K.; Megias, M.; Niewczas, K.; Pickering, L.; Sobczyk, S.; Thompson, L; Wret, C.}

\address{%
$^{1}$ \quad CERN European Organization for Nuclear Research, CH-1211 Genéve 23, Switzerland\\
$^{2}$ \quad University of Sheffield, Department of Physics and Astronomy, Sheffield, United Kingdom\\
$^{3}$ \quad IRFU, CEA, Université Paris-Saclay, F-91191 Gif-sur-Yvette, France\\
$^{4}$ \quad University of Tokyo, Institute for Cosmic Ray Research, Kamioka Observatory, Kamioka, Japan\\
$^{5}$ \quad University of Rochester, Department of Physics and Astronomy, Rochester, New York, U.S.A.\\
$^{6}$ \quad Departamento de Física Atómica, Molecular y Nuclear, Universidad de Sevilla, 41080 Sevilla, Spain\\
$^{7}$ \quad Wroclaw University, Faculty of Physics and Astronomy, Wroclaw, Poland\\
$^{8}$ \quad Ghent University, Department of Physics and Astronomy, Gent, Belgium\\
$^{9}$ \quad Royal Holloway University of London, Department of Physics, Egham, Surrey, United Kingdom}

\abstract{The NEUT event generator is a widely-used tool to simulate neutrino interactions for energies between 10s of MeV and a few TeV. NEUT plays a crucial role in neutrino oscillation analyses for the T2K and Hyper-K experiments, providing the primary simulation of the neutrino interactions whose final-state products are measured to infer the oscillation parameters. NEUT is also capable of simulating nucleon decay and hadron scattering. These proceedings
present an expansion of NEUT to simulate electron scattering before showing comparisons to experimental measurements and using discrepancies to derive an empirical correction to NEUT's treatment of nuclear removal energy.}

\begin{document}


\section{Introduction}
\label{sec:intro}

Robust predictions of neutrino interaction cross sections, alongside reliable uncertainties on them, are a critical ingredient in neutrino oscillation measurements at GeV-scale energies~\cite{Alvarez-Ruso:2017oui}. The currently-operating T2K~\cite{T2K:2011qtm} and Super-Kamiokande~\cite{Super-Kamiokande:2002weg} experiments, as well as the future Hyper-K~\cite{Hyper-Kamiokande:2018ofw} experiment, use the NEUT event generator to provide such predictions. A comprehensive overview of NEUT can be found in Ref.~\cite{Hayato:2021heg}. NEUT is able to simulate neutrino interactions with nuclei from the threshold of quasi-elastic scattering ($\sim$1-50 MeV) up to TeV energy scales using a variety of theory-driven interaction models that can be chosen and configured by the user. Hadronic final-state interactions within the nuclear medium are modelled using an intranuclear cascade, which is tuned thanks to NEUT's ability to predict hadron-scattering measurements~\cite{PinzonGuerra:2018rju}. NEUT has a rich history, having been developed in the 1980s for atmospheric neutrino and nucleon decay studies and being used for Super-Kamiokande's Nobel prize winning analysis of neutrino oscillations~\cite{Super-Kamiokande:1998kpq}. NEUT continues to be developed primarily by the collaborators of experiments that use it.

In these proceedings, a first step towards expanding NEUT's repertoire of scattering simulations to include electron-nucleus interactions is presented. The comparison to electron-scattering measurements is a valuable tool to validate any neutrino-interaction model. The form factors describing the vector component of neutrino interactions can be directly inferred from electron-scattering experiments, and the nuclear target in the weak and electromagnetic interaction is identical. Electron-scattering experiments are also typically much higher precision than their neutrino scattering counterparts, being able to produce mono-energetic beams with interaction cross sections typically more than eight orders of magnitude larger (for the GeV-scale energies relevant to neutrino oscillation experiments). It is therefore unsurprising that recent years have seen a plethora of analyses of electron-scattering data tailored towards constraining poorly understood neutrino interaction physics~\cite{CLAS:2021neh,electronsforneutrinos:2020tbf,Ankowski:2020qbe}. 

\section{Electron scattering in NEUT}
\label{sec:weaktoem}

Weak neutral current (NC) quasi-elastic (QE) neutrino-nucleus interactions are used as a basis to construct electromagnetic (EM) QE pseudo electron scattering prediction within NEUT. This is to remove any possible issues with the pure isovector contributions of charged-current (CC) QE interactions. However, using NC interactions as a framework assumes a neutral lepton in both the initial and final state. Neither of these are subject to the effects of the Coulomb potential that an electron would experience in a scattering interaction, and so a shift in final-state lepton energy is made to correct for this. In this analysis we use 3.6 MeV for carbon and 4.1 MeV for oxygen~\cite{Gueye:1999mm,Bodek:2018lmc}.

In addition to the Coulomb correction, it is also necessary to change form factors used for NCQE scattering to replace the weak couplings with the EM couplings. These alterations are two-fold: the first is to set the axial form factors, $F_A$ and $F_P$, to zero, and second is to change the electric and magnetic form factors to the simpler form required by the EM interactions. Finally, the weak coupling constant is replaced by its EM counterpart. 

NEUT offers several nuclear models to predict the lepton scattering on a nuclear target. This analysis uses the Benhar spectral function model, which is based on the work presented in Ref.~\cite{Hayato:2021heg,Benhar:1994hw}, and is used as the primary input model for T2K's neutrino oscillation analyses. This approach relies on the plane wave impulse approximation (PWIA) to factorise the QE cross-section calculation into an expression containing a single-nucleon factor alongside a ``spectral function’' (SF). The single nucleon factor is altered from the usual neutrino scattering case using the prescription presented above, whilst the SF component remains unchanged. The SF is a two-dimensional distribution describing the probability of finding a nucleon with some particular momentum, $|\mathbf{p}|$, and removal energy, $E_{rmv}$, which corresponds to the energy required to remove the nucleon from the nuclear potential. 

Whilst the SF model has been relatively successful in describing electron and neutrino scattering measurements, the use of a factorised PWIA calculation means that at low momentum transfer ($q_3$ less than $\sim$ 300-400 MeV~\cite{Rocco:2016ejr}), the model predictions are expected to be missing important components. For example, the use of PWIA implies that the nucleon removal energy ``seen'' by the incoming lepton is independent of the four-momentum of the impinging particle. More sophisticated models that consider effects beyond PWIA predict a relationship: the distortion of the outgoing nucleon wave function by the nuclear potential in a relativistic mean field (RMF) model finds a linear correlation between the momentum transfer and a shift from the PWIA expected peak of the QE interaction as a function of energy transfer, $q_0$~\cite{Gonzalez-Jimenez:2014eqa, Megias:2017PhD}. Such shifts in the QE peak are characteristic of a shifted removal energy~\cite{Bodek:2018lmc}. Theory-driven corrections to the SF model have been suggested to account for this~\cite{Ankowski:2014yfa} but are not included in this study.

As detailed in Ref.~\cite{Ankowski:2014yfa}, a mismodelling of shifts in the QE peak position can cause important biases on neutrino experiments' ability to reconstruct neutrino energy, which causes biases on measurements of neutrino oscillations. It is therefore imperative that potential mismodelling of such shifts is scrutinised using both theory and experiment. Neutrino experiments have attempted to constrain this~\cite{T2K:2021xwb,MINERvA:2019ope} but the broad incoming neutrino energy spectrum makes such analyses very challenging. Electron-scattering measurements are much better suited for this, and are routinely used to study QE peak position alterations from expectation (see e.g. ~\cite{Bodek:2018lmc,Ankowski:2014yfa}).

\section{Comparison to inclusive electron-scattering data}
\label{sec:comptodata}

\begin{figure}[htb]
    \centering
    \begin{subfigure}[b]{0.49\textwidth}
    \captionsetup{justification=centering}
        \centering 
        \includegraphics[width=\textwidth]{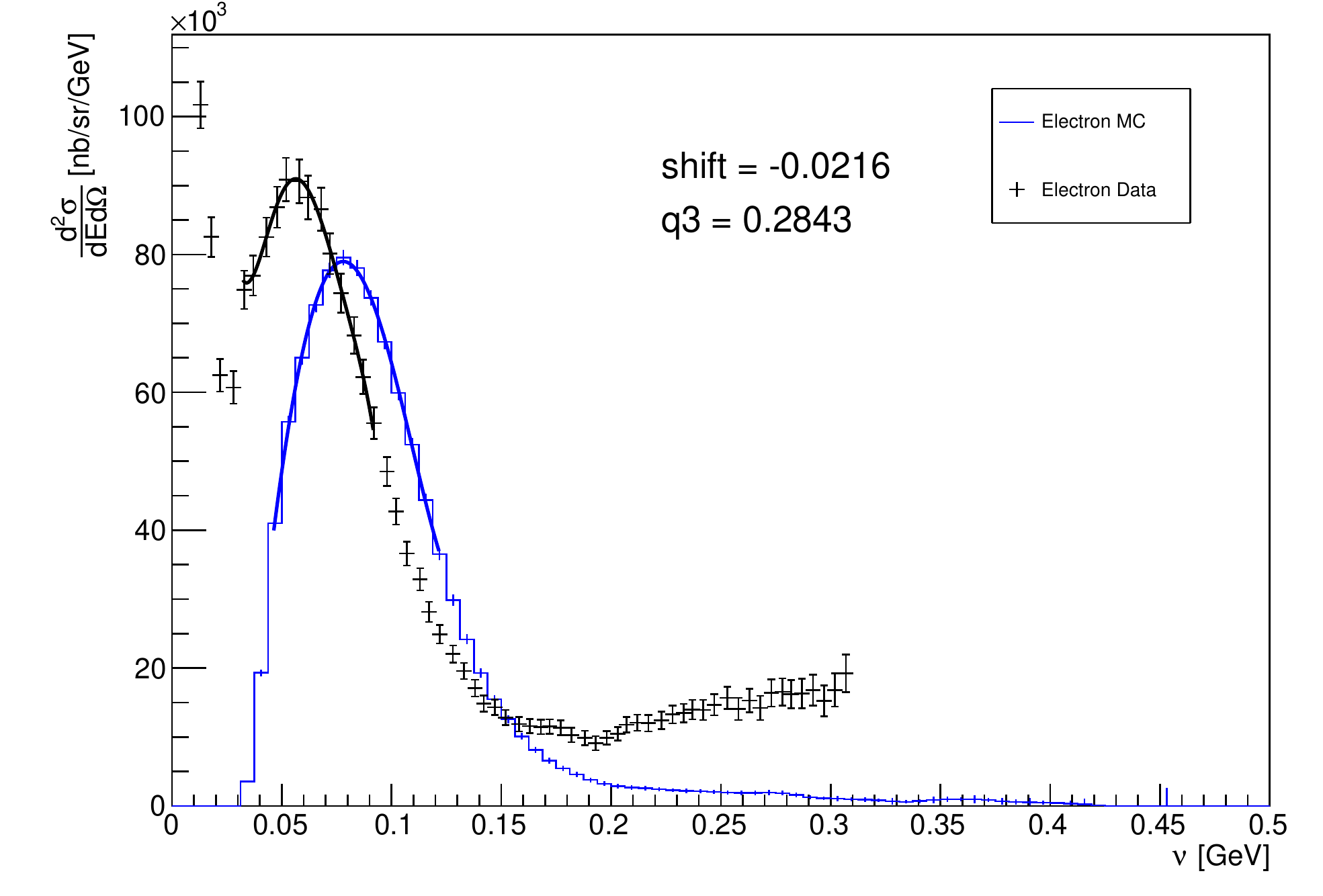}
        \caption{$E=480~\textrm{MeV}, \theta=36^\circ$}
    \end{subfigure}
    \begin{subfigure}[b]{0.49\textwidth}
    \captionsetup{justification=centering}
        \centering 
        \includegraphics[width=\textwidth]{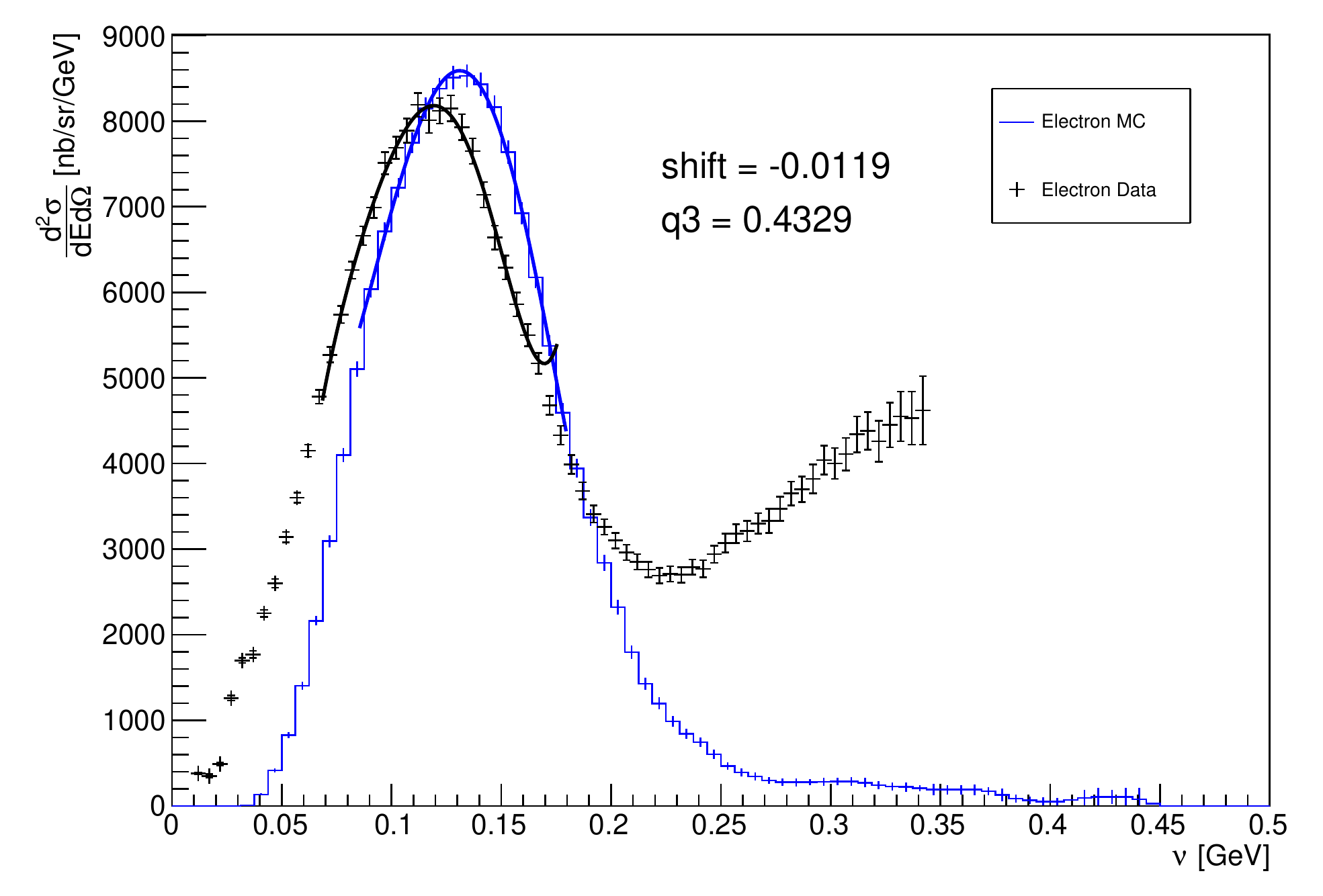}
        \caption{$E=480~\textrm{MeV}, \theta=60^\circ$}
    \end{subfigure}
    \vspace{2mm}

    \begin{subfigure}[b]{0.49\textwidth}
    \captionsetup{justification=centering}
        \centering 
        \includegraphics[width=\textwidth]{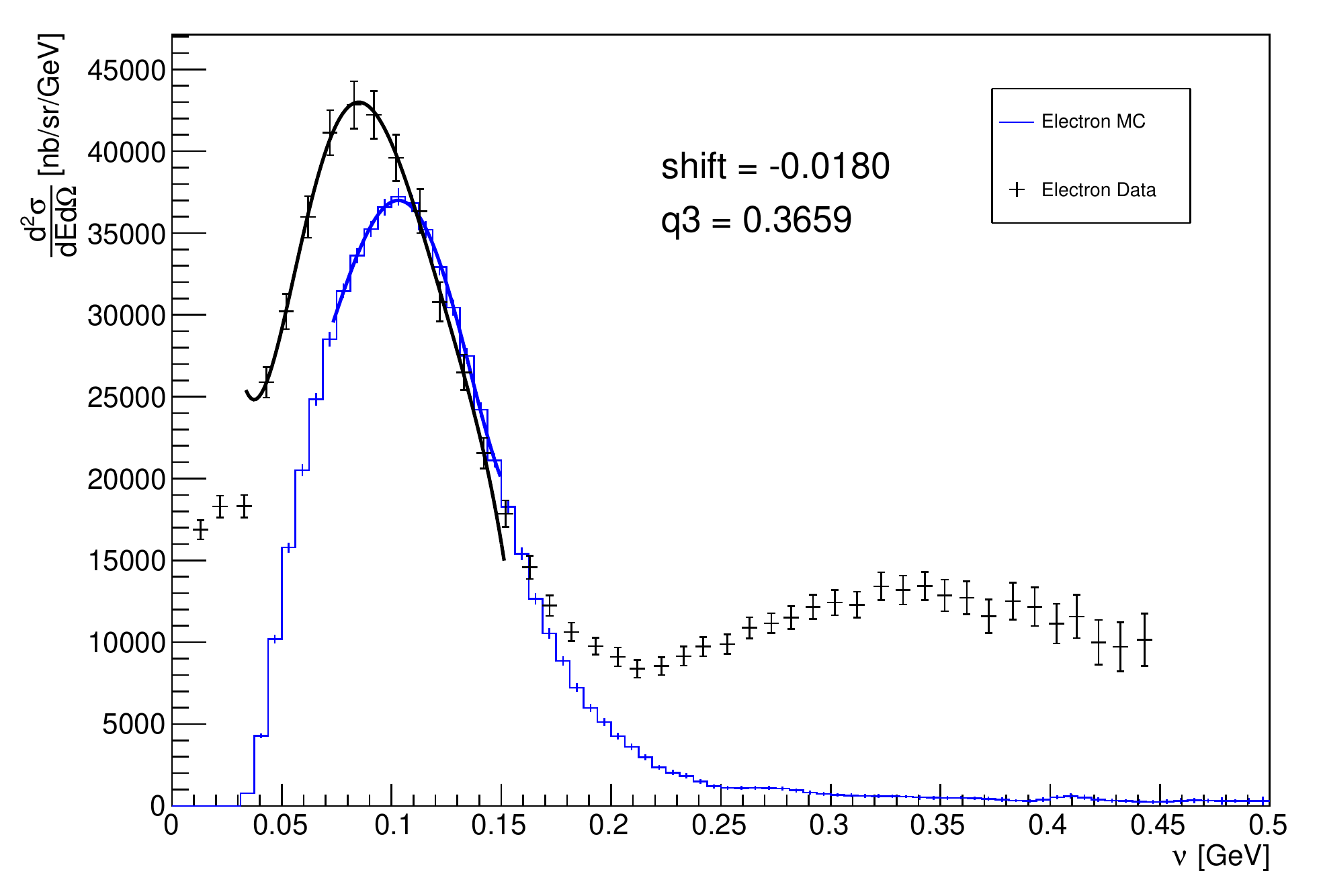}
        \caption{$E=620~\textrm{MeV}, \theta=36^\circ$}
    \end{subfigure}
    \begin{subfigure}[b]{0.49\textwidth}
    \captionsetup{justification=centering}
        \centering 
        \includegraphics[width=\textwidth]{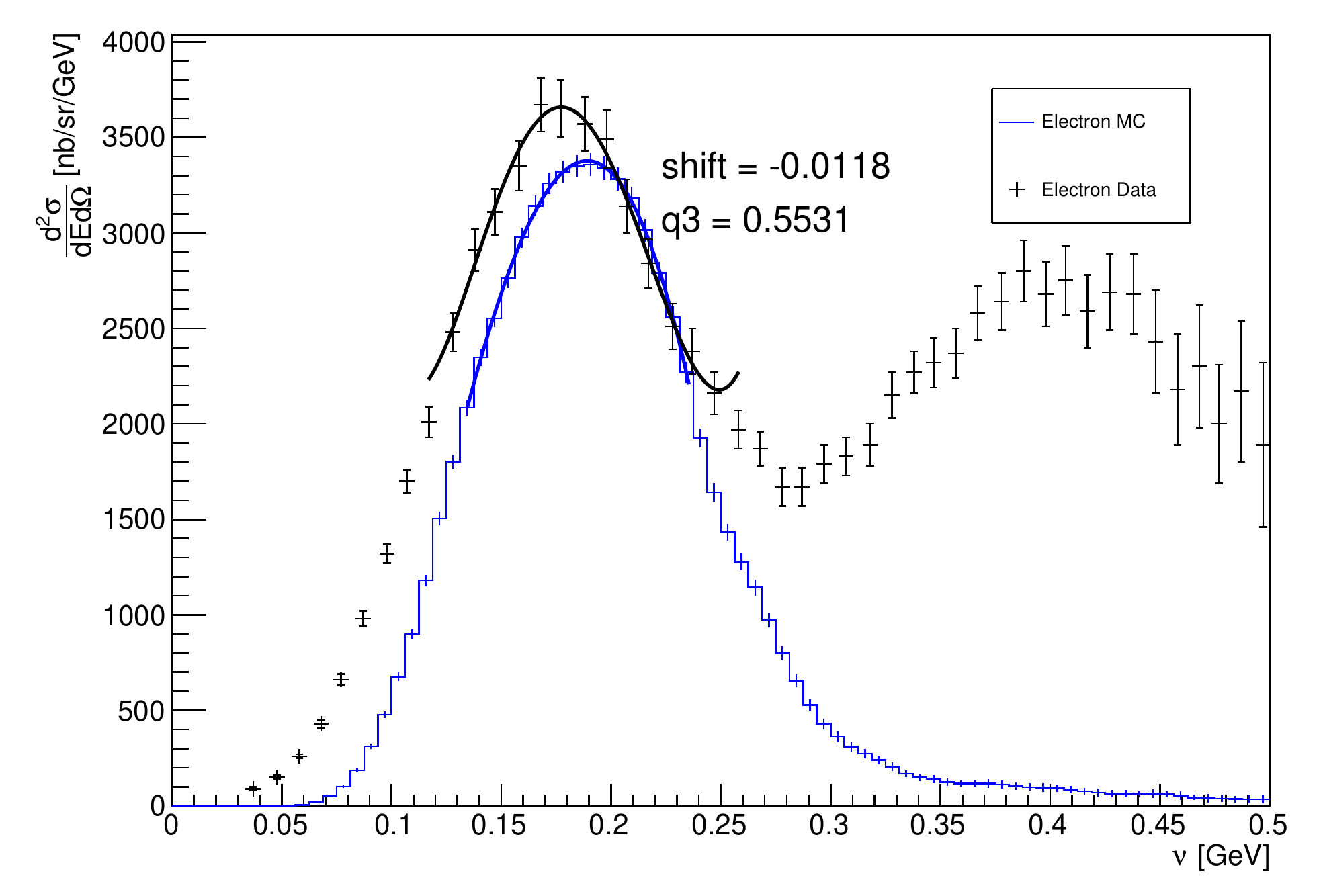}
        \caption{$E=620~\textrm{MeV}, \theta=60^\circ$}
    \end{subfigure}
    \vspace{2mm}
    
    \caption{Inclusive electron cross-section measurements~\cite{qearchive} (data points) compared to quasi-elastic electron-scattering generation in NEUT (blue histogram) as a function of energy transfer. The comparison is made for a number of different scattering angles and beam energies. The peaks are fitted with a fifth-order polynomial to extract the position (the smooth lines) and the difference between the peak positions is shown on each plot (``shift'', in GeV). The $q_3$ calculated at the peak position of the cross-section measurement (``$q_3$'', in GeV/c).} 
    \label{fig:escat_fit}
\end{figure}

Inclusive electron cross-section measurements on carbon~\cite{qearchive} as a function of the incident beam energy, outgoing lepton angle, and the measured energy transfer, are compared to NEUT EMQE predictions in Fig.~\ref{fig:escat_fit}. A complete description of the data throughout the energy transfer range would require the prediction of inelastic interaction channels, which is beyond the scope of this work. However, in the region of the QE peak, a qualitative comparison between NEUT and the measurements can be made. Within this region, the NEUT prediction is shifted with respect to the data. The size of the shift is derived by fitting a fifth-order polynomial to determine the peak position of the prediction and the data, which is shown as smooth lines in Fig.\ref{fig:escat_fit}. Measurements with expected large inelastic contributions overlapping with QE peak region are not used in this analysis, thereby limiting the range of momentum transfers considered. 

\section{Determining a momentum-transfer dependent correction}
\label{sec:SFq3dep_analysis}

\begin{figure}[htbp]
\centering
\begin{subfigure}[b]{0.49\textwidth}
    \captionsetup{justification=centering}
    \includegraphics[width=\textwidth]{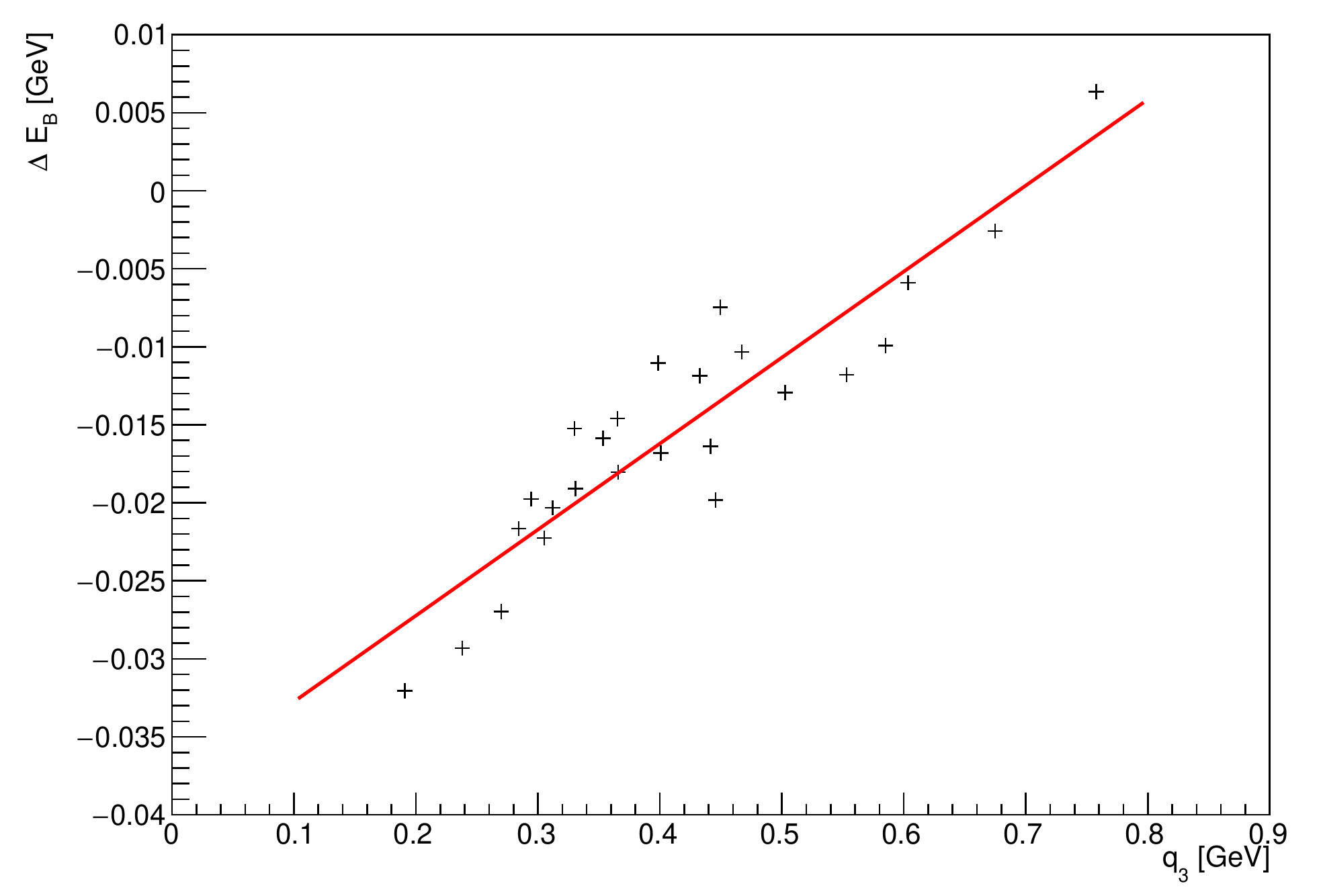}
    \caption{Before correction}
\label{fig:q3depshift}
\end{subfigure}
\begin{subfigure}[b]{0.49\textwidth}
    \captionsetup{justification=centering}
    \includegraphics[width=\textwidth]{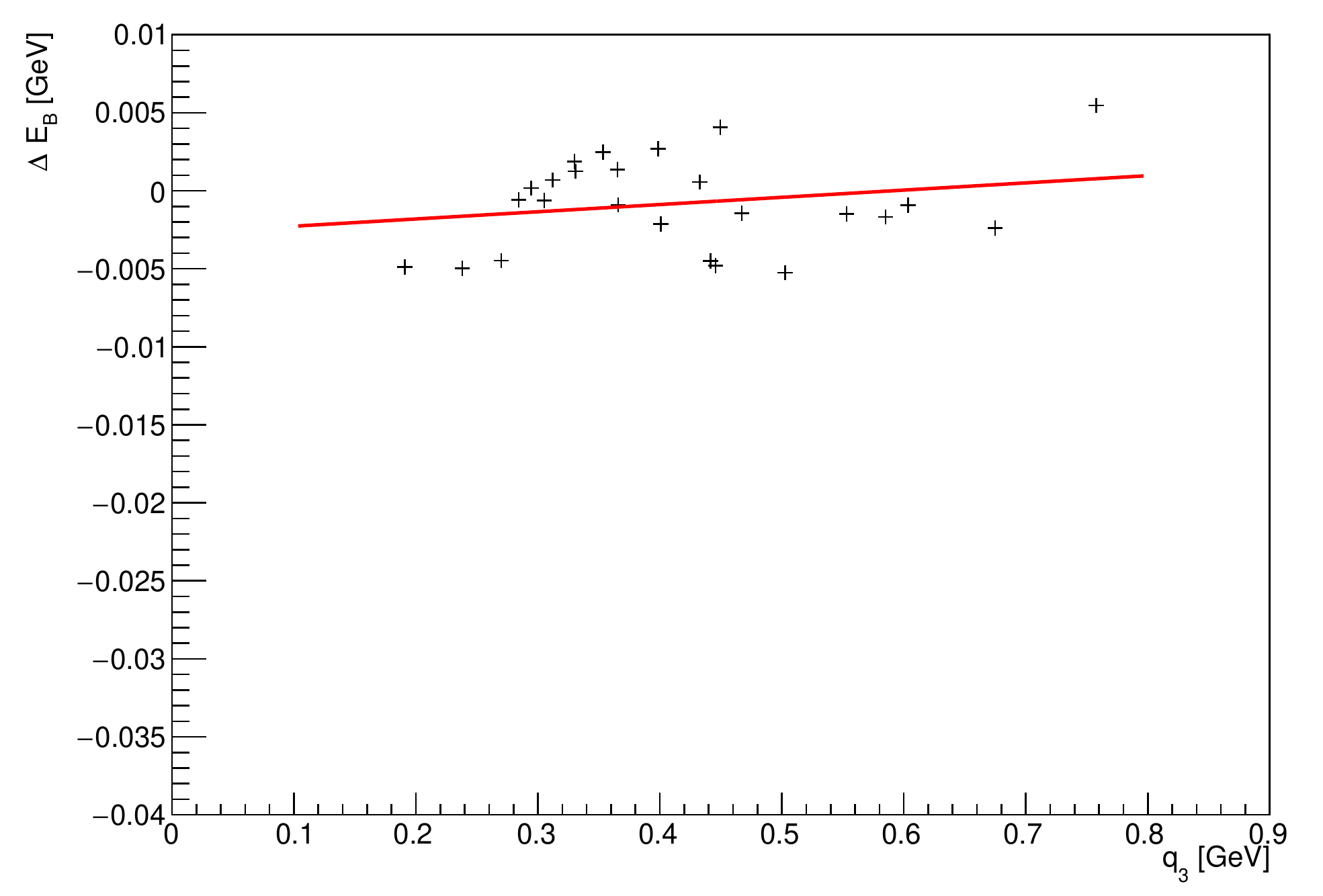}
    \caption{After correction}
\label{fig:q3depshift_post}
\end{subfigure}
\caption{The shifts derived from Fig.~\ref{fig:escat_fit} are shown as a function of momentum transfer to reveal an approximately linear relationship, as described by the fitted line (left). The derived shift is interpreted as a $E_{rmv}$ correction and applied within NEUT, and the shifts are re-derived for the corrected model, showing there is no need for further corrections (right).}
\label{fig:q3SFshift}
\end{figure}

As discussed in Sec.~\ref{sec:weaktoem}, given the limitations of NEUT's PWIA-based nuclear model, the shift with respect to the QE peak is not surprising and an evolution as a function of the momentum transfer is expected. To evaluate this, the derived shift is plotted against the momentum transfer (taken at the measured peak) in the left panel of Fig.~\ref{fig:q3SFshift}. A linear fit, which is also shown in the figure, is qualitatively reasonable to describe the evolution of the shift. The gradient was found to be 0.056, which corresponds to the same evolution as predicted from the aforementioned RMF model~\cite{Gonzalez-Jimenez:2014eqa,Megias:2017PhD}.

A naive correction for the breaking of factorisation can be built by shifting the removal energy of NEUT's SF model as a function of the momentum transfer according to the linear fit. As this is an empirical correction to the model, it is able to shift some small fraction of the SF removal energy distribution to be negative at large values of momentum transfer. To alleviate this issue, negative removal energies are truncated at zero. It was verified that the truncation has a small impact on the corrected NEUT predictions. 

Applying the correction during event generation has the expected result of more accurately predicting the electron-scattering data. This is demonstrated in the right panel of Fig.~\ref{fig:q3SFshift}, which shows that the evolution derived from the corrected distribution is close to zero for all values of momentum transfer.

\section{Relevance for neutrino scattering}

\begin{figure}[htb]
    \centering
    \vspace{-3mm}
    \begin{subfigure}[b]{0.5\textwidth}
        \centering \includegraphics[width=\textwidth]{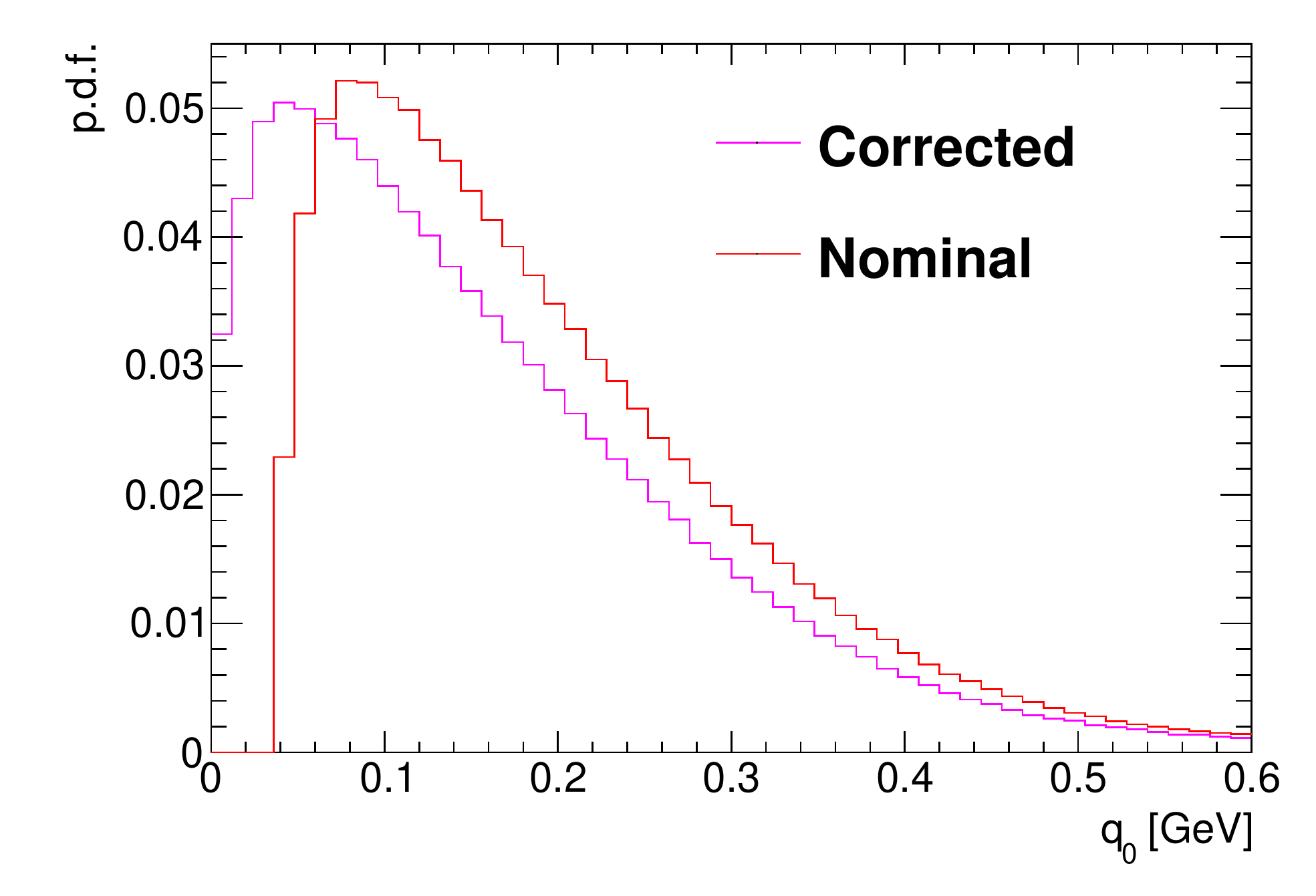}
    \end{subfigure}\hfill
    \begin{subfigure}[b]{0.5\textwidth}
        \centering \includegraphics[width=\textwidth]{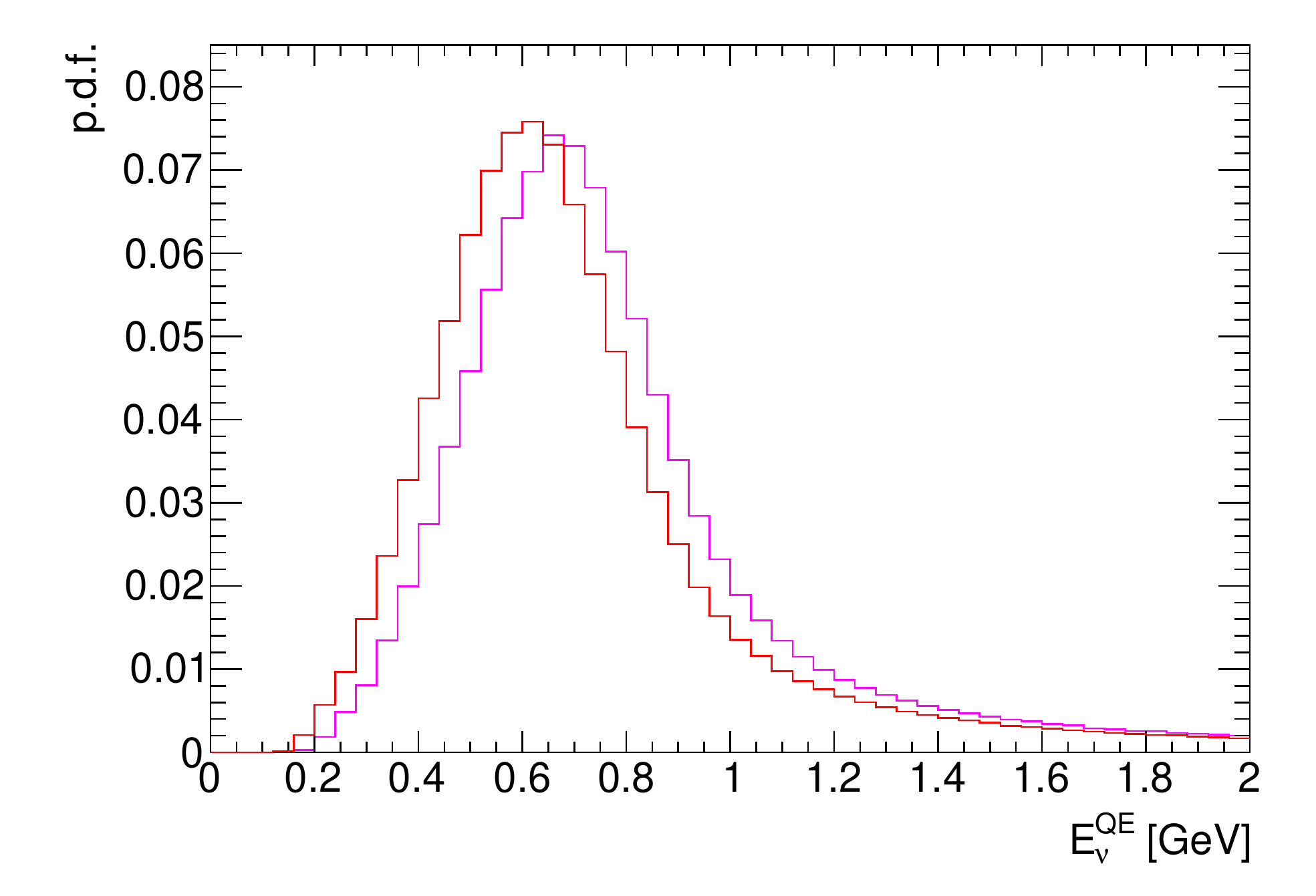}
    \end{subfigure}
    
    \begin{subfigure}[b]{0.5\textwidth}
        \centering \includegraphics[width=\textwidth]{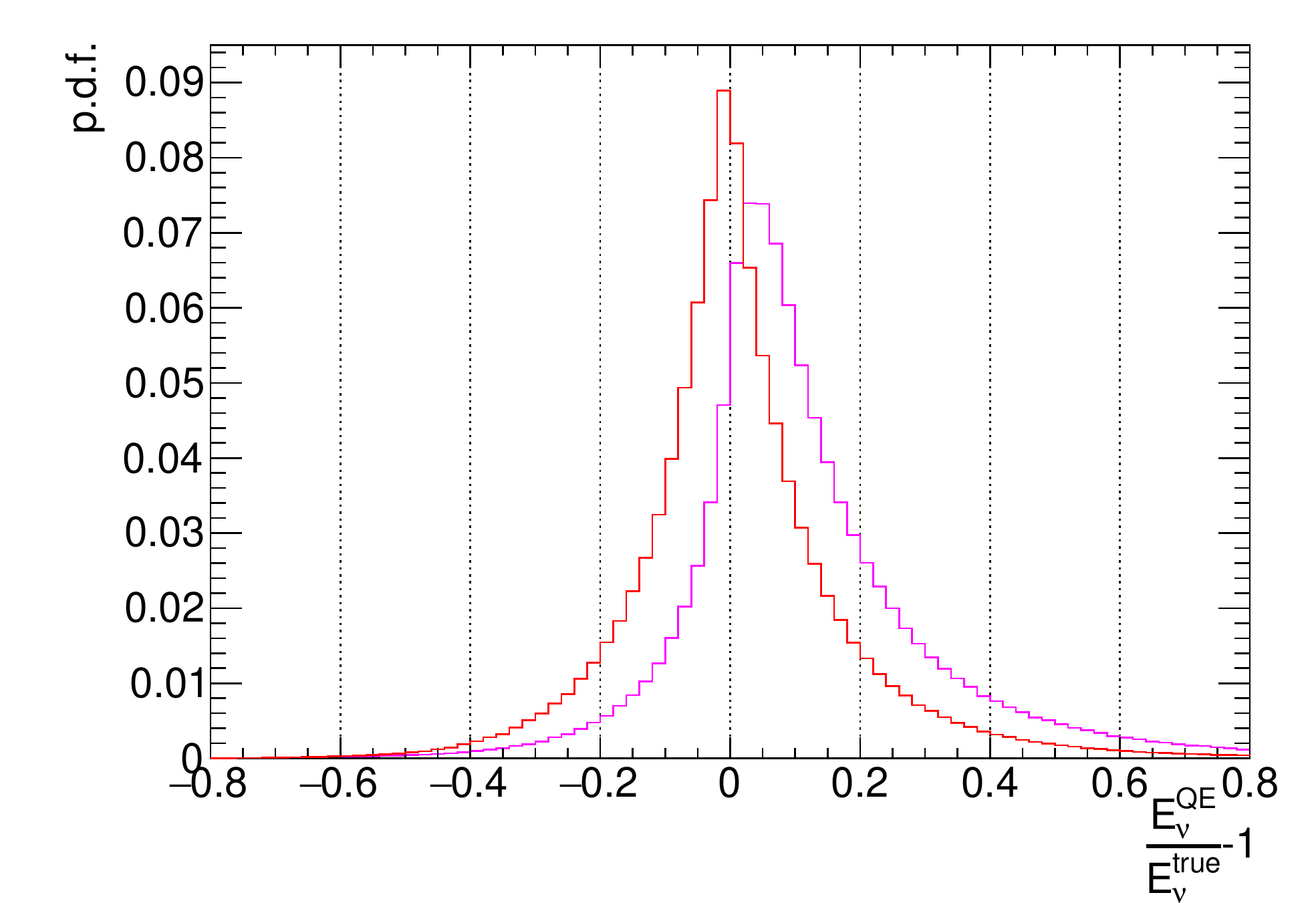}
    \end{subfigure}\hfill
    \begin{subfigure}[b]{0.5\textwidth}
        \centering \includegraphics[width=\textwidth]{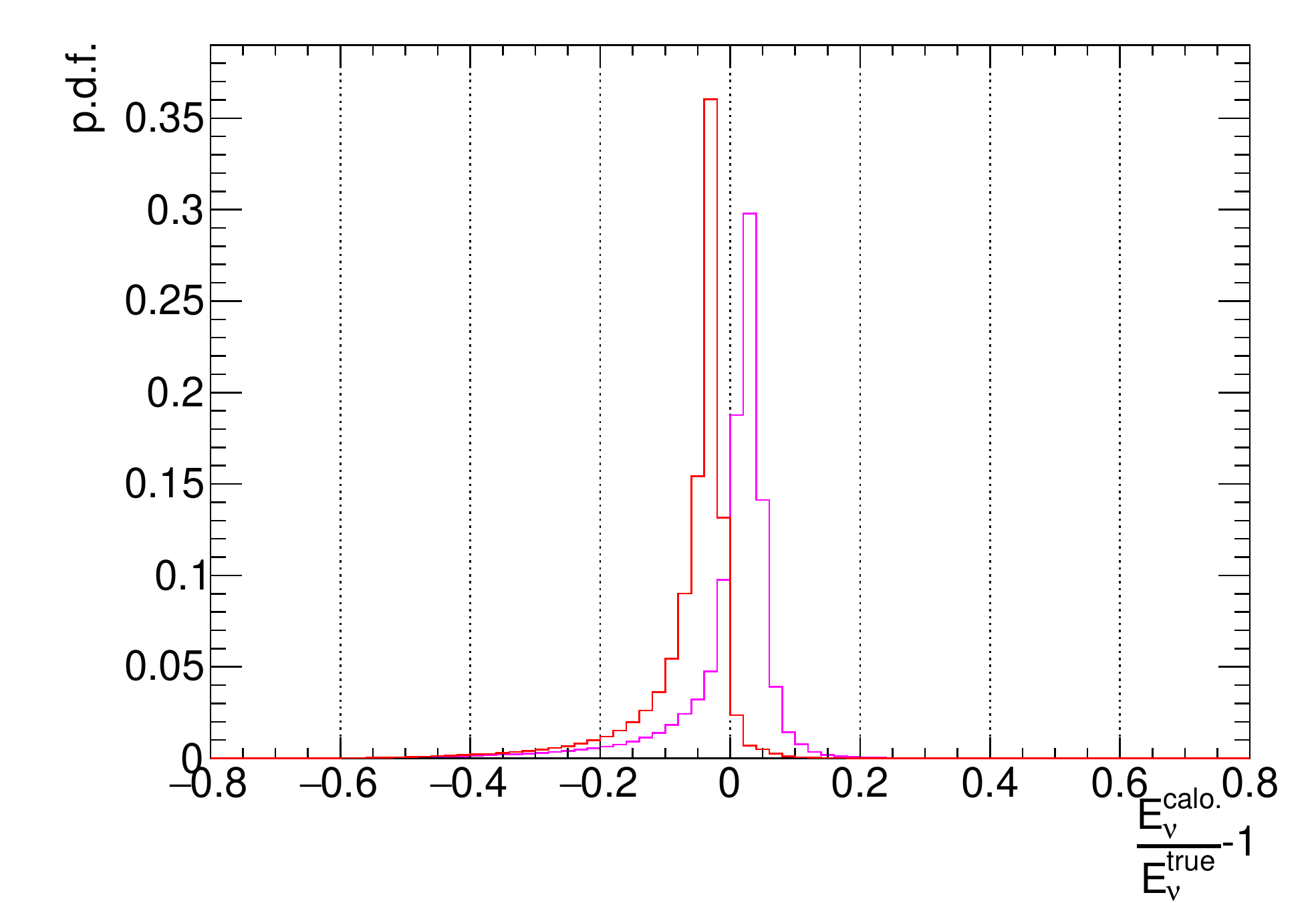}
    \end{subfigure}
\caption{The NEUT simulation for CCQE interactions applying (magenta) and not applying (red) the derived $q_3$-dependent removal energy correction on a carbon target using the T2K $\nu_\mu$ flux. The upper plots show the cross section shape as a function of the energy transfer (left) and reconstructed neutrino energy (right), using the neutrino energy estimator from T2K (using lepton kinematics only~\cite{T2K:2021xwb}). The lower plots show the bias and spread of the neutrino energy estimators used by T2K (left) and NOvA (right, using total energy deposits estimated using $E_{avail}$ from Ref.~\cite{MINERvA:2015ydy}).}
\label{fig:impactfornus}
\end{figure}

It is interesting to consider the effect of the correction on NEUT's simulation of neutrino interactions. To study this, the impact of the correction on neutrino energy reconstruction metrics and CCQE cross sections on a carbon target using the T2K $\nu_\mu$ flux are shown in Fig.~\ref{fig:impactfornus}. The correction causes a shift and broadening of the cross section as a function of energy transfer and reconstructed neutrino energy. This is expected from Fig.~\ref{fig:q3SFshift}, as for most regions of momentum transfer the correction tends to shift the removal energy to smaller values, thereby permitting interactions with lower energy transfer and biasing of the neutrino energy reconstruction.
The impact on the neutrino energy reconstruction bias is considerable and, if taken as an unconstrained uncertainty, may be a dominant effect. This applies equally whether neutrino energy is reconstructed using T2K's approach, based on only the outgoing lepton kinematics~\cite{T2K:2021xwb}, or with the NOvA's approach, using total observed energy deposits~\cite{NOvA:2020rbg}. For this reason, T2K's latest oscillation analysis~\cite{t2k_neutrino2022} uses a model which fully brackets this uncertainty~\cite{nuinttalk}.

Whilst the impact of the correction on neutrino energy reconstruction spread and bias appears to be large, the extrapolation of the correction from electron-scattering data is not guaranteed to be reliable. The qualitative nature of the analysis and the need to truncate the correction to avoid non-physical removal energies, discussed in Sec~\ref{sec:SFq3dep_analysis}, is important to keep in mind. Furthermore, the $q_3$ range covered by the considered electron-scattering measurements is situated below the expected peak $q_3$ for T2K (0.5 GeV, extending up to about 1 GeV), meaning much of the correction is applied to NEUT events in a $q_3$ range which is relatively poorly constrained. Moreover, whilst the nuclear target is unchanged, the aspects of the nuclear response probed by EM and weak interactions are not guaranteed to be the same. 

\section{Conclusions}

The NEUT event generator has been updated to simulate EMQE interactions. Predictions from the updated generator have been compared to inclusive, QE dominated, electron-scattering measurements to derive an empirical momentum-transfer dependent correction to the nuclear removal energy in NEUT's SF nuclear model. The correction has an important impact on neutrino energy reconstruction metrics, and is considerably relevant for current and future neutrino oscillation analyses. A more sophisticated analysis involving the simulation of inelastic scattering channels would permit a more robust correction to be derived. A detailed comparison of electron and neutrino cross-section predictions from the corrected NEUT to more sophisticated nuclear models, which go beyond the PWIA prescription, would better inform its applicability to neutrino-nucleus interaction simulations for neutrino oscillation experiments. 



\begin{adjustwidth}{-\extralength}{0cm}

\reftitle{References}


\bibliography{bibNoTitle.bib}

\end{adjustwidth}
\end{document}